\documentclass{egpubl_pg_short}
\usepackage{pg2017s}

 \WsPaper         
 \electronicVersion 

\ifpdf \usepackage[pdftex]{graphicx} \pdfcompresslevel=9
\else \usepackage[dvips]{graphicx} \fi

\PrintedOrElectronic

\usepackage{t1enc,dfadobe}

\usepackage{egweblnk}
\usepackage{cite}

\usepackage{amsmath}
\usepackage{bm}
\usepackage{array}
\usepackage{listings}
\usepackage[noend]{algpseudocode}
\usepackage{algorithm}
\usepackage{siunitx}
\usepackage{stackengine}
\usepackage{subfigure}

\algblockdefx[Loop]{Loop}{EndLoop}[1][]{\textbf{Loop} #1}{\textbf{End Loop}}

\usepackage{caption} 
\usepackage{enumerate}
\usepackage[T1]{fontenc}

\lstdefinestyle{bottom}{
  float,
  floatplacement=b
}

\title{Efficient barycentric point sampling on meshes}

\author[J. Portsmouth]
{\parbox{\textwidth}{
\centering J. Portsmouth \\
\centering Advanced Micro Devices
}}

\begin{document}

\maketitle

\begin{abstract}
We present an easy-to-implement and efficient analytical inversion algorithm for the unbiased random sampling of a set of points on a triangle mesh whose surface density is specified by barycentric interpolation of non-negative per-vertex weights. The correctness of the inversion algorithm is verified via statistical tests, and we show that it is faster on average than rejection sampling.  
\begin{classification} 
\CCScat{Computer Graphics}{G.3}{Probability and Statistics}{Probabilistic algorithms}
\end{classification}

\end{abstract}

\section{Related Work}
\label{sec:relatedwork}

Point sampling on meshed surfaces is useful in a variety of computer graphics contexts \cite{Gross:2007:PG:1202384}. 
Here we focus on the relatively simple problem of sampling a spatially inhomogeneous, unbiased random distribution of points on a mesh given a prescribed point density per unit area varying over the mesh, which has received surprisingly little attention. 

There has been much work on the generation of point distributions with desired local statistical characteristics for applications such as point stippling and blue noise sampling \cite{Corsini:2012:EFS:2197070.2197094,Cline:2009:DTS:2383586.2383611}, however these techniques are complex to implement, generally not very efficient, and deliberately introduce bias into the sampling.

Statistically unbiased inhomogeneous random sampling is clearly useful in a variety of contexts, for example in Monte Carlo sampling. For the homogeneous case, there is a standard inversion algorithm for sampling a random point on a triangle mesh with uniform probability \cite{Pharr:2010:PBR:1854996, Turk:1990:GRP:90767.90772, smith2004sampling}. 
For the inhomogeneous case,
\emph{rejection sampling} \cite{Press:1992:NRC:148286} provides a general and relatively efficient solution, as was noted (for example) by Yan et. al \cite{6811174}, however as a random algorithm this presents efficiency problems. Sik et. al \cite{Sik:2013:FRS} improved on rejection sampling by subdividing the mesh until each triangle can be considered to have uniform density without losing too much accuracy, then applying the homogeneous inversion algorithm. However this requires a relatively complex preprocessing stage.

Arvo et. al \cite{10.1109/RT.2007.4342601,SS2M} extended the inversion algorithm to deal with a density which varies linearly within each face of a triangle mesh (corresponding to barycentric interpolation). However they expressed their solution in the form of a set of cubic equations, without simplifying further or developing a practical, tested algorithm. In this paper we revisit the approach of Arvo et al., and provide a more explicit algorithm than previously, with statistical validation and performance comparison to rejection sampling. 

\section{Method}

Consider the general problem of independently sampling random points on a three-dimensional triangle mesh, where the probability per unit area $p_X(\mathbf{x})$ of choosing a given point $\mathbf{x}$ is proportional to a non-negative scalar weight on the mesh, $\phi(\mathbf{x})$.
This weight can be interpreted as specifying the relative surface density (i.e. number per unit area) of sampled points. 
Normalization then implies:
\begin{equation}
p_X(\mathbf{x}) = \frac{\phi(\mathbf{x})} {\sum_i \int_{T_i} \phi(\mathbf{y}) \,\mathrm{d}A_i(\mathbf{y})} \nonumber
\end{equation}
where $\mathrm{d}A_i(\mathbf{y})$ is the area element on the three-dimensional surface of triangle $T_i$.
This may be factored into the discrete probability of choosing a given triangle, multiplied by the conditional PDF (with area measure) of choosing a point within that triangle:
\begin{eqnarray}
p_X(\mathbf{x}) &=& p_{T}(T_i) \; p_{X|T}(\mathbf{x} \vert T_i) \nonumber \\
&=& \frac{\int_{T_i} \phi(\mathbf{y}) \,\mathrm{d}A_i(\mathbf{y})}  {\sum_{k} \int_{T_k} \phi(\mathbf{y})\,\mathrm{d}A_k(\mathbf{y})} \cdot
    \frac{\phi(\mathbf{x})}            {\int_{T_i} \phi(\mathbf{y}) \,\mathrm{d}A_i(\mathbf{y})} \ . \nonumber
\end{eqnarray}
Defining the area-weighted average of $\phi$ over triangle $T_i$ (with area $A_i$) as $\langle\phi\rangle_i \equiv \int_{T_i} \phi(\mathbf{x}) \,\mathrm{d}A_i(\mathbf{x}) / A_i$, then the discrete probability of each triangle can be written as
\begin{eqnarray}
 p_{T}(T_i) &=& \frac{A_i\langle\phi\rangle_i}{\sum_{k} A_k \langle\phi\rangle_k } \ , \nonumber
\end{eqnarray}
and the conditional area-measure PDF of a point $\mathbf{x}$ in a given triangle $T_i$ is 
\begin{equation} \label{conditional_pdf_formula}
	p_{X|T}(\mathbf{x} \vert T_i) = \frac{\phi(\mathbf{x})}{A_i \langle\phi\rangle_i} \ .
\end{equation}
Sampling from the discrete probability distribution $p_{T}(T_i)$ to choose a triangle is done via the CDF
\begin{equation}
P_{T}(T_i) = \frac{\sum_{j\le i} A_j\langle\phi\rangle_j}{\sum_{k} A_k \langle\phi\rangle_k } \ , \nonumber
\end{equation}
i.e. sample a uniform random deviate (a random variable drawn from the uniform distribution on the open interval $[0,1)$), then find index $k$ such that $P_{T}(T_{k-1}) \le \xi < P_{T}(T_k)$ (with $P_{T}(T_k)=0$ by convention for $k<k_{\mathrm{min}}$). This is usually done via bisection search, though note that more efficient algorithms exist such as that described by Sik et al. \cite{Sik:2013:FRS}.
%
%
It remains to sample from the conditional PDF $p_{X|T}(\mathbf{x} \vert T_i)$.

A completely general method for sampling from $p_{X|T}(\mathbf{x} \vert T_i)$  given any function $\phi(\mathbf{x})$ is provided by \emph{rejection sampling}. 
We first find $\phi_{\mathrm{max},i} = \max(\phi(\mathbf{x}): \mathbf{x}\in T_i)$ (this can be precomputed). We then choose a random point in the triangle drawn from a uniform distribution. This is most easily done by parameterizing points $\mathbf{x}\in T_i$ as $\mathbf{x} = u\,\mathbf{v}^i_{u} + v\,\mathbf{v}^i_{v} + w\,\mathbf{v}^i_{w}$, where $\mathbf{v}^i_{u}$, $\mathbf{v}^i_{v}$, $\mathbf{v}^i_{w}$ are the triangle vertices and the per-triangle barycentric coordinates $(u, v, w)$ are each in the range $[0,1]$ with $u+v+w=1$. Then uniform random sampling on the triangle is done via the formulas \cite{Pharr:2010:PBR:1854996}
\begin{eqnarray} \label{uniform_triangle_sample}
u = 1 - \sqrt{\xi_1} \ , \quad v = (1 - u) \,\xi_2 \nonumber
\end{eqnarray}
where $(\xi_1, \xi_2)$ are uniform random deviates.
We then decide to accept this trial sample by drawing another uniform random deviate $\xi_3$ and testing whether $\xi_3 \,\phi_{\mathrm{max},i} < \phi(\mathbf{x})$. If this is true, we accept $\mathbf{x}$ as the sample, otherwise we draw another trial sample of $\mathbf{x}$ and continue, until acceptance.
While the rejection sampling method is very general, as a random algorithm it does not provide any strict guarantees about the number of random samples which will be taken. 

However, in the common case of a weight defined by per-vertex barycentric weighting, the inversion method provides an analytical formula for the sampled points\cite{10.1109/RT.2007.4342601,SS2M}. Per-vertex weighting means we associate with each of the three vertices of triangle $T_i$ a non-negative real weight. Let us denote the weights of the three vertices $\mathbf{v}^i_{u}$, $\mathbf{v}^i_{v}$, $\mathbf{v}^i_{w}$ as $\phi_u$, $\phi_v$, $\phi_w$ respectively. Assuming the triangle is not degenerate, we may express the barycentric coordinates in terms of position: $u(\mathbf{x})$, $v(\mathbf{x})$, $w(\mathbf{x})$. 
Barycentric interpolation then defines the weight at all points $\mathbf{x}\in T_i$ via (equivalent to linear interpolation within the triangle):
\begin{equation}
\phi(\mathbf{x}) = u(\mathbf{x})\,(\mathbf{\phi}_{u}-\mathbf{\phi}_{w}) + v(\mathbf{x}) \,(\mathbf{\phi}_{v}-\mathbf{\phi}_{w}) + \mathbf{\phi}_{w} \ . \nonumber
\end{equation}
This is a common scheme for interpolating per-vertex weights to produce a $C^0$ continuous function on the mesh. 
Integrals over the triangle area elements $\mathrm{d}A_i$ may be completed by a change of variables to barycoordinates, via the standard identity $\mathrm{d}A_i = 2A_i \, \mathrm{d}u \,\mathrm{d}v$.  
It follows that the area-averaged barycentric weight is equal to the mean vertex weight:
\begin{equation}
\langle\phi\rangle_i = \frac{\phi_u + \phi_v + \phi_w}{3} \ . \nonumber
\end{equation}
Using Eqn.~(\ref{conditional_pdf_formula}), the normalized conditional PDF in barycentric coordinate measure is then given by
\begin{equation} \label{pdf_uv}
	p_{U,V}(u, v) = p_{X|T}(\mathbf{x} \vert T_i) \frac{\mathrm{d}A_i}{\mathrm{d}u \,\mathrm{d}v} = 2 \frac{\phi(u, v)}{\langle\phi\rangle_i} \ .
\end{equation}

\begin{figure}[!t] 
  \centering
  \includegraphics[width=4.6cm, angle=-90]{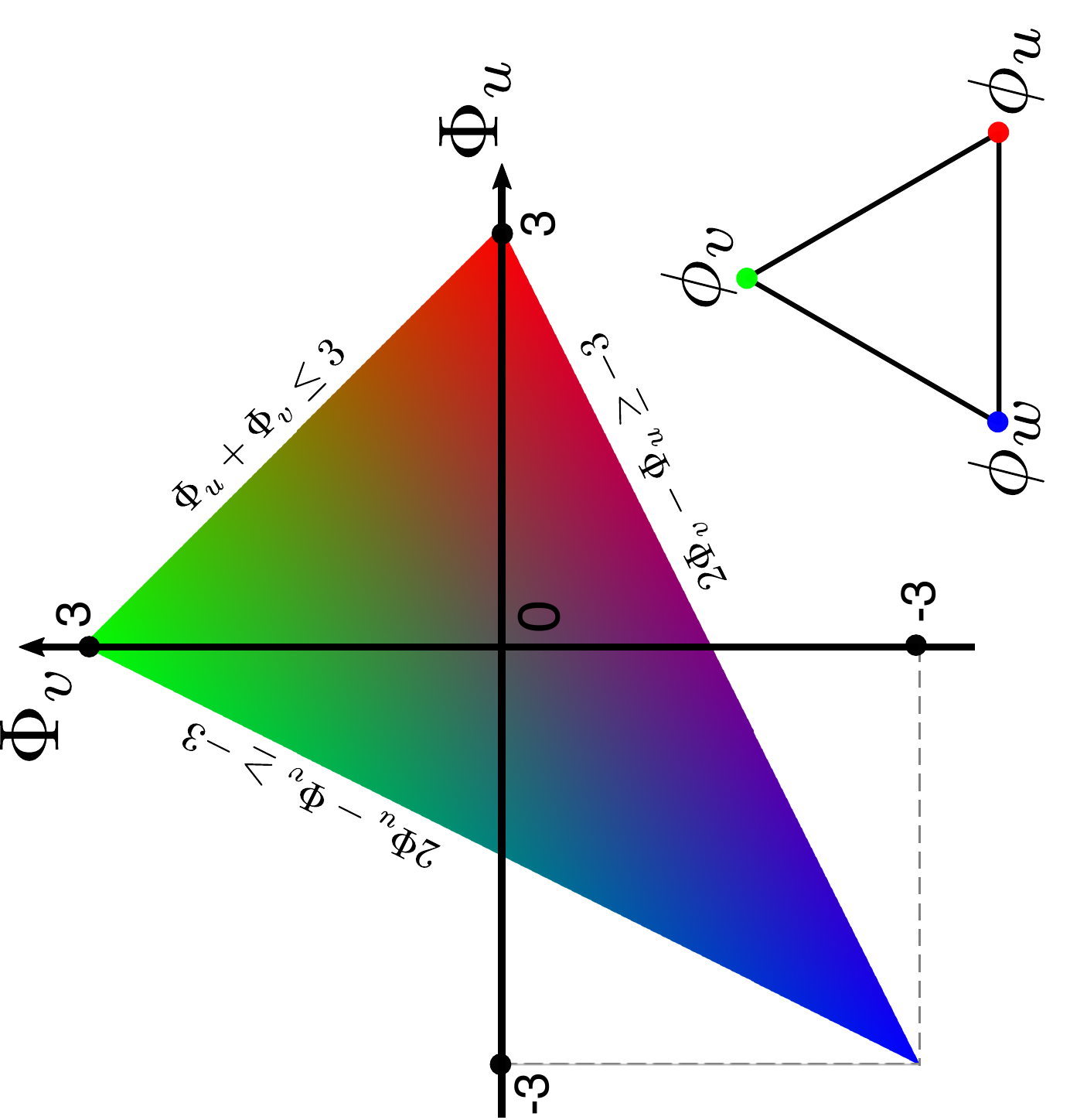}
  \caption{All possible non-negative triangle vertex-weights $\phi_u$ (red), $\phi_v$ (green), and $\phi_w$ (blue) map into the triangle shown in the plane of relative weights $(\Phi_u, \Phi_v)$. The PDF $p_{U,V}(u, v)$ associated with the triangle is a function only of its position in this plane, given by Eqn.~(\ref{pdf_uv_solution}).
  \label{weight_triangle}}
\end{figure}

We now introduce the normalized \emph{relative weights}
\begin{eqnarray}
\Phi_u &\equiv& \frac{\phi_u-\phi_w} {\langle\phi\rangle_i} \ , \quad \Phi_v \equiv \frac{\phi_v-\phi_w} {\langle\phi\rangle_i} \ . \nonumber
\end{eqnarray}
Each possible set of relative weights maps into a point in the $(\Phi_u, \Phi_v)$ plane (i.e. each such point represents all triangles which have the same relative weights). Expressing the problem in terms of this two-dimensional vector of weights leads to some simplification in the analytical expressions compared to Arvo's approach \cite{SS2M}. Note that the case where $\langle\phi\rangle_i=0$, i.e. all zero weights, can be ignored as the probability of sampling a point in such a zero-weighted triangle is zero. It follows from the non-negativity of the weights that $(\Phi_u, \Phi_v)$ satisfy various inequalities
and thus lie within the triangular region in the $(\Phi_u, \Phi_v)$ plane depicted in Figure~\ref{weight_triangle}. The uniform weighting corresponds to the origin $(\Phi_u, \Phi_v)= (0,0)$.  
\begin{lstlisting}[style=bottom, caption={Routine to sample the $u$ barycoordinate.}, label={lst:U_NEWTON}, float]
double <@\textbf{\detokenize{U}}@>( double Phi_u, double Phi_v, const double tol = 5.0e-3 )
{
    double r = <@\textcolor{red}{RAND}@>(); double l = (2.0*Phi_u - Phi_v)/3.0;
    const int maxIter = 20; double u = 0.5; int n=0;
    while (n++ < maxIter) {
        double u1 = 1.0-u;
        double P  = u*(2.0-u) - l*u*u1*u1 - r;
        double Pd = max(u1*(2.0 + l*(3.0*u-1.0)), DBL_EPSILON);
        double du = max(min(P/Pd, 0.25), -0.25); u -= du;
        u = max(min(u, 1.0-DBL_EPSILON), DBL_EPSILON);
        if (fabs(du) < tol) break;
    }
    return u;
}
\end{lstlisting}
In these coordinates, the barycentric-measure PDF Eqn.~(\ref{pdf_uv}) reduces to
\begin{equation} \label{pdf_uv_solution}
	p_{U,V}(u, v) = 2 \left( u\,\Phi_u + v\,\Phi_v + 1 - \frac{\Phi_u + \Phi_v}{3} \right) \ . \nonumber
\end{equation}

We now describe how to sample points from this PDF.
Integration gives the marginal PDF of $u$:
\begin{eqnarray}
	p_U(u) &=& \int_0^{1-u} p_{U,V}(u, v) \; \mathrm{d}v \ . \nonumber
\end{eqnarray}
The cumulative density function (CDF) for the marginal PDF of $u$, $\int_0^{u} p_U(u') \,\mathrm{d}u'$ is given by (and similarly for the CDF of $v$, $P_V(v)$)
\begin{eqnarray} \label{cdf1}
\!\!\!\!P_U(u) \!&=&\!  u\left(2-u\right) - \frac{\left(2\Phi_u - \Phi_v\right)}{3} u(u-1)^2	\ . 
\end{eqnarray}
Inversion of $P_U(u)=\xi_u$ where $\xi_u$ is a uniform random deviate yields the sampled value of $u$. It is efficient to solve this by Newton's method via the update rule: 
\begin{equation}
u_{n+1} = u_n - \frac{P_U(u_n)-\xi_u}{P'_U(u_n)} \ . \nonumber
\end{equation}
An example implementation of this sampling routine in C is provided in Listing~\ref{lst:U_NEWTON} (where \textcolor{red}{\texttt{RAND}} generates a uniform random deviate). This includes tolerances to keep the solution within bounds, and limits the step size to $1/2$ to aid convergence (as statistically verified in Section \ref{sec:validation}).
\begin{figure}[!t] 
  \centering
  \includegraphics[width=3.9cm, angle=-90]{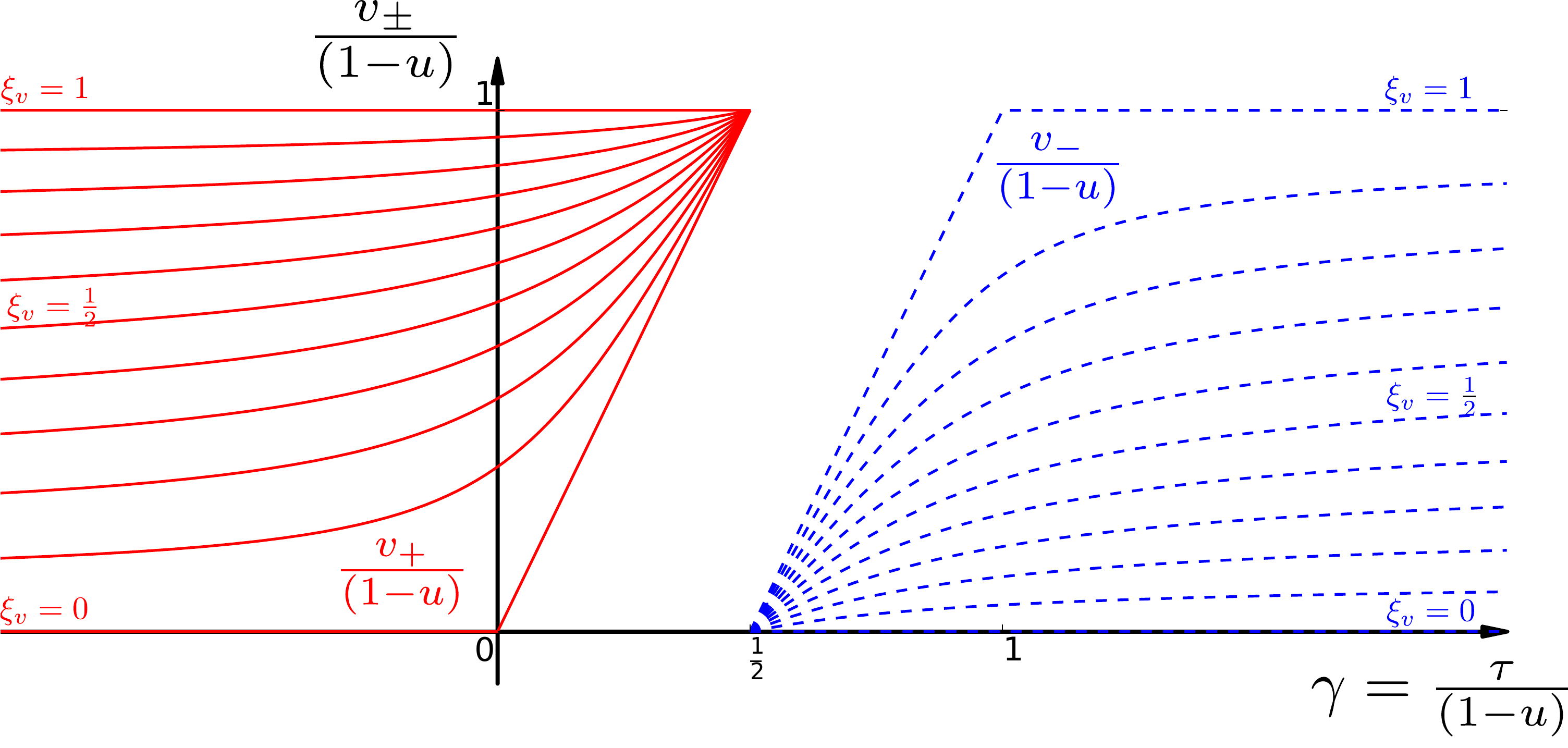}
  \caption{Illustrates the family of solutions for the barycoordinate $v$ (Eqn.~(\ref{vquadsolve})), given the previously sampled barycoordinate $u$ and random deviate $\xi_v$, as a function of $\tau(u, \Phi_u, \Phi_v)$ defined in Eqn.~(\ref{define_tau}). \label{v_solutions}}
\end{figure}

In order to sample $v$, we require the CDF for $v$ conditional on the sampled value of $u$, given by 
\begin{equation}
P_V(v \vert u) = \int_0^{v} p_{V|U}(v' \vert u) \,\mathrm{d}v' = \frac{1}{p_U(u)} \int_0^{v} p_{U,V}(u, v') \,\mathrm{d}v' \ . \nonumber
\end{equation}
Evaluating this gives
\begin{equation} 
P_V(v \vert u) = \frac{2v}{p_U(u)} \left[ 1 + \left(u-\frac{1}{3}\right)\,\Phi_u + 
  \left(\frac{v}{2}-\frac{1}{3}\right)\,\Phi_v \right] \ . \nonumber
\end{equation}
Inversion of $P_V(v \vert u)=\xi_v$ (where $\xi_v$ is again a uniform random deviate) gives the sampled value of $v$. As $P_V(v \vert u)$ is quadratic in $v$, this inversion has two solutions
\begin{equation} \label{vquadsolve}
 v_{\pm} = \tau \pm \sqrt{ \,\tau^2 (1-\xi_v) + \left(\tau + u - 1\right)^2\xi_v }
\end{equation}               
where the square root here denotes the principal square root, and 
\begin{eqnarray} \label{define_tau}
\tau(u, \Phi_u, \Phi_v) &\equiv& \frac{1}{3} - \frac{1+\left(u-\frac{1}{3}\right)\,\Phi_u}{\Phi_v} \ .
\end{eqnarray} 
Here $\tau$ ranges over the entire real line, i.e. $\tau \in [-\infty, \infty]$.
As $\Phi_v$ approaches zero (uniform weighting), $\tau$ diverges; however, in this limit the inversion can be simplified to:
\begin{eqnarray} 
v \rightarrow (1-u) \, \xi_v  + O(\left|\tau\right|^{-1}) \quad \mathrm{as} \quad \left|\Phi_v\right| \rightarrow 0, \left|\tau\right| \rightarrow \infty \ . \nonumber
\end{eqnarray} 

If $\tau\gg 0$, then clearly the $v_-$ branch must be chosen (in order that $v \in [0,1])$. Similarly, if $\tau\ll 0$, then the $v_+$ branch must be chosen. Figure~\ref{v_solutions} shows how the solution for $v$ varies as a function of the random deviate $\xi_v$ and $\tau$. 
From this figure it is intuitively clear that for general $\tau$, the correct choice of branch is given by
\begin{equation} \label{signchoice}
v = \left\{
\begin{array}{l l}
\!v_+ & \mbox{if } \; \tau \le (1-u)/2   \ , \\
\!v_- & \mbox{if } \; \tau > (1-u)/2   \ .
\end{array}  \right.
\end{equation}
\begin{lstlisting}[style=bottom, caption={Routine to sample the $v$ barycoordinate.}, label={lst:V}, float]
double <@\textbf{V}@>( double u, double Phi_u, double Phi_v )
{
	double r = <@\textcolor{red}{RAND}@>();
	const double epsilon = 1.0e-6 
	if (fabs(Phi_v) < epsilon) return (1.0 - u)*r;
	double tau = 1.0/3.0 - (1.0 + (u-1.0/3.0)*Phi_u)/Phi_v;
	double tmp = tau + u - 1.0;
	double q = sqrt(tau*tau*(1.0-r) + tmp*tmp*r);
	return tau <= 0.5*(1.0-u) ? tau + q : tau - q;
}
\end{lstlisting}
\begin{proof}
Let $\tau = \gamma \left(1 - u\right)$, which implies:
\begin{equation}
\frac{v_{\pm}}{(1-u)} = \gamma \pm \sqrt{\gamma^2 + \xi_v(1-2\gamma)} \ . \nonumber
\end{equation}
First consider the case $\gamma>1/2$. Then the term inside the square root satisfies the following inequality
\begin{equation}
\gamma^2 + \xi_v(1-2\gamma) > \left|\gamma-1\right|^2 \nonumber
\end{equation}
since $(1-2\gamma)<0$ and $\xi_v<1$. Thus the following inequalities are satisfied:
\begin{equation}
\frac{v_+}{(1-u)} > \gamma + \left|\gamma-1\right| > 1 \ , \quad 
\frac{v_-}{(1-u)} < \gamma - \left|\gamma-1\right| < 1 \ . \nonumber
\end{equation}
Therefore since $v\le(1-u)$, if $\gamma>1/2$ the $v_-$ branch must be chosen. The analogous argument for $\gamma<1/2$ shows that the $v_+$ branch must be chosen in that case.
\end{proof}
\begin{algorithm}[!t]
  \caption{Inversion sampling (per-vertex barycentric)}\label{inversion}
  \begin{algorithmic}[1]
  	\Function{INVERSION}{$T_i$}\Comment{Samples from $p_{X|T}(\mathbf{x} \vert T_i)$}
	\State Sample uniform random deviates $\xi_u$, $\xi_v$
	\State Compute $u$ by solve of $P_U(u)=\xi_u$ (Eqn.~(\ref{cdf1}), Listing~\ref{lst:U_NEWTON})
    \State Compute $v$ (Eqn.~(\ref{signchoice}), Listing~\ref{lst:V})
    \State \textbf{return} $\mathbf{x}(u, v)$
   \EndFunction
  \end{algorithmic}
\end{algorithm}
Algorithm~\ref{inversion} summarises the resulting method for inversion sampling $(u, v)$.
A C implementation for sampling $v$ via this method is provided in routine \textbf{V} (Listing~\ref{lst:V}).

\begin{figure*}[!tb]
\captionsetup{width=0.9\textwidth}
\hspace*{\fill}
\centering
\hfill
\stackunder[5pt]{\includegraphics[width=0.3\linewidth]{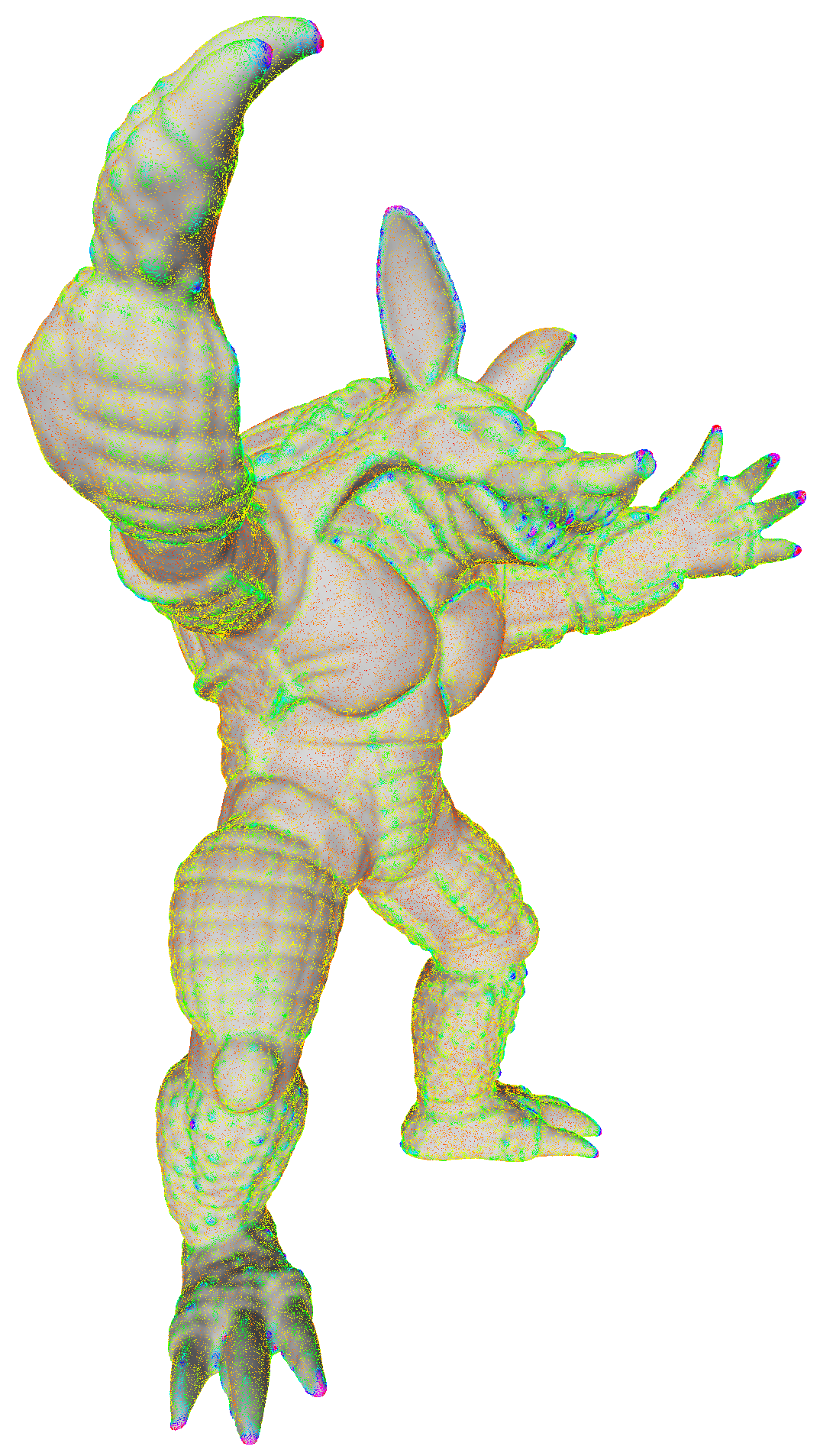}}{}
\hfill
\stackunder[5pt]{\includegraphics[width=0.333\linewidth]{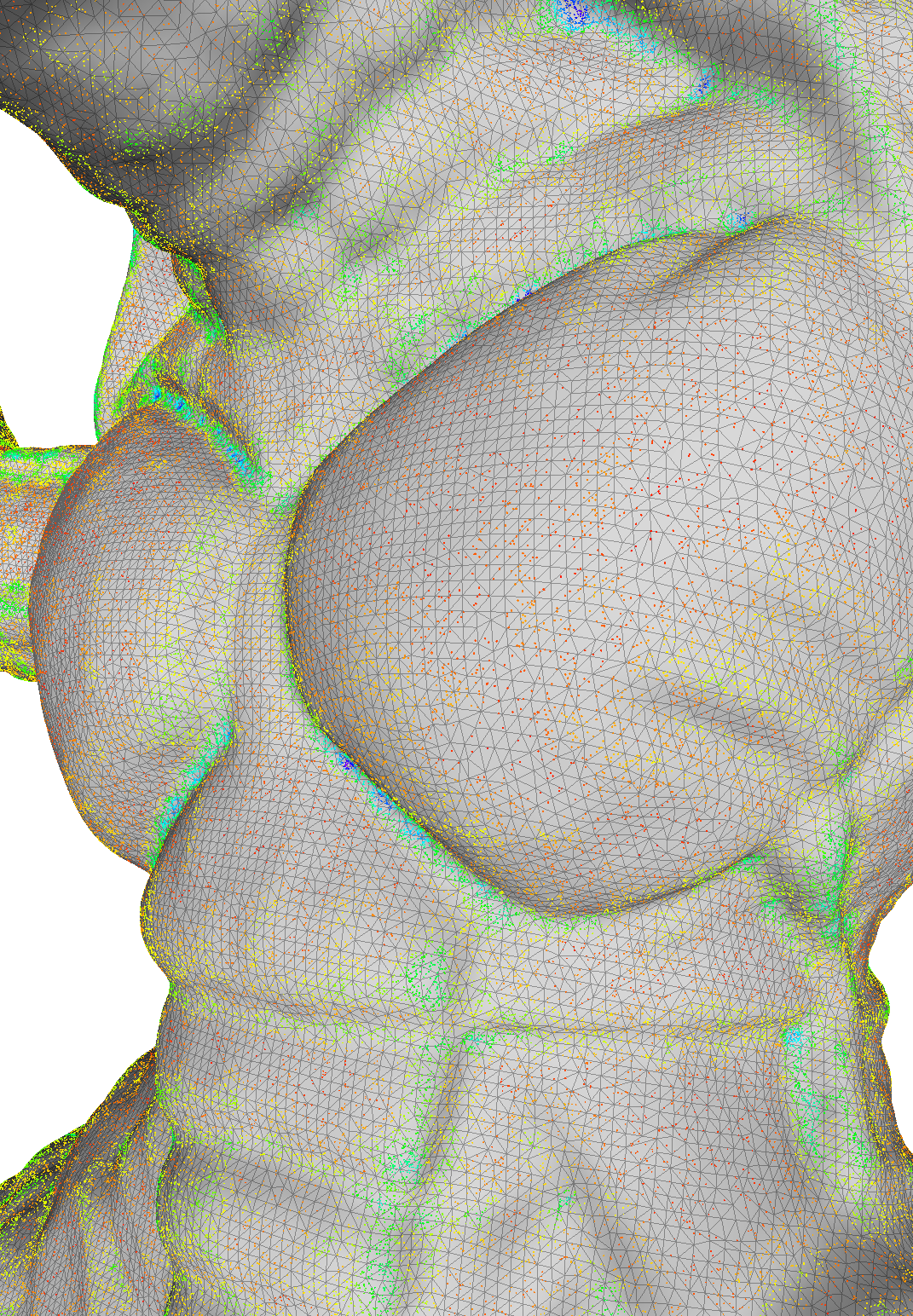}} 
\hfill
\caption{One million points distributed on a mesh with density proportional to per-vertex curvature magnitude, via inversion sampling. The point hue indicates the variation of curvature magnitude from minimum to maximum. The detail on the right shows the relation of the sampled points to the underlying mesh.}
\label{fig:points_curv}
\end{figure*}

\begin{figure*}[!tb]
\captionsetup{width=0.9\textwidth}
\hspace*{\fill}
\centering
\hfill
\stackunder[5pt]{\includegraphics[width=0.333\linewidth]{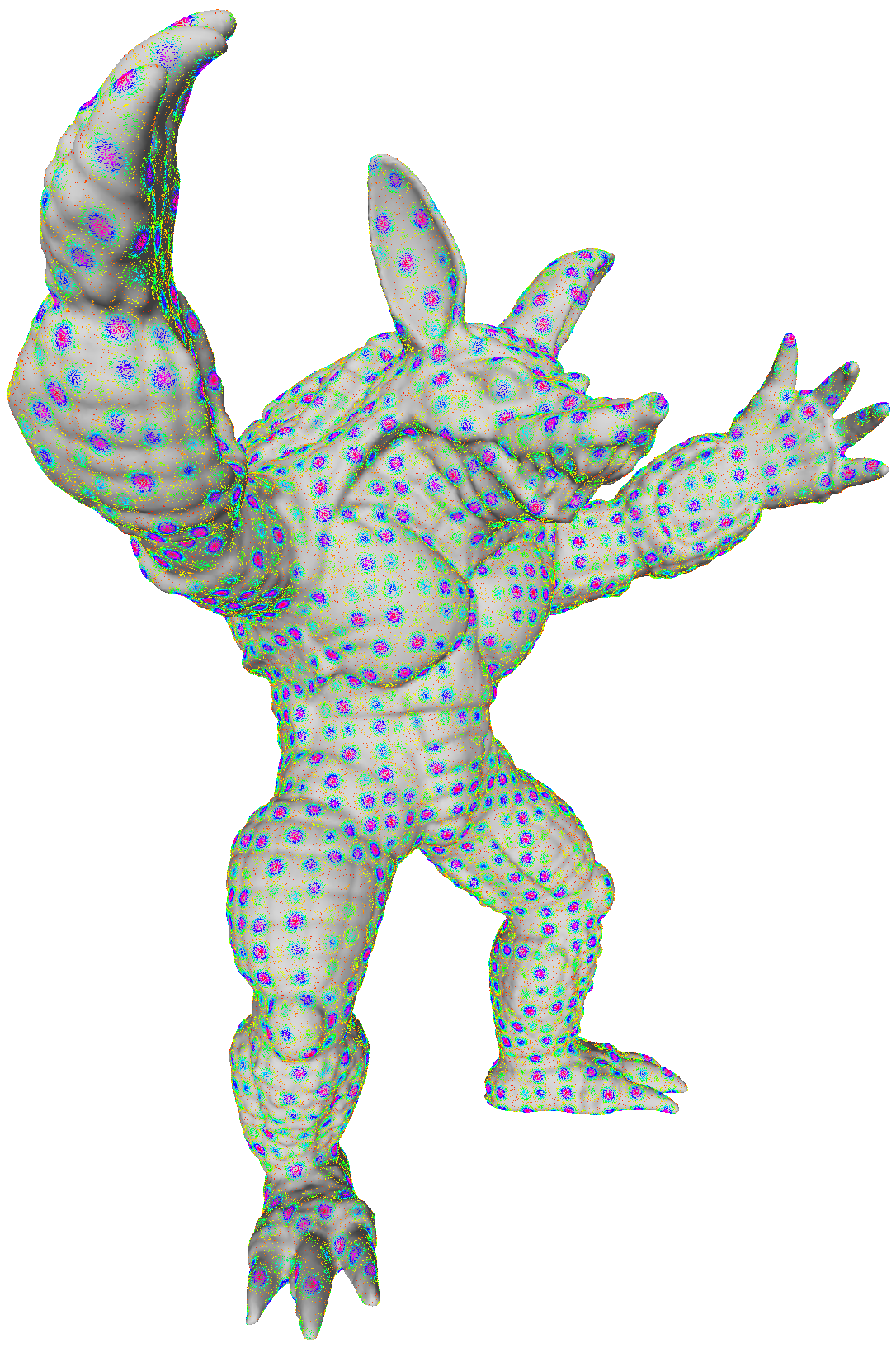}}{}
\hfill
\stackunder[5pt]{\includegraphics[width=0.333\linewidth]{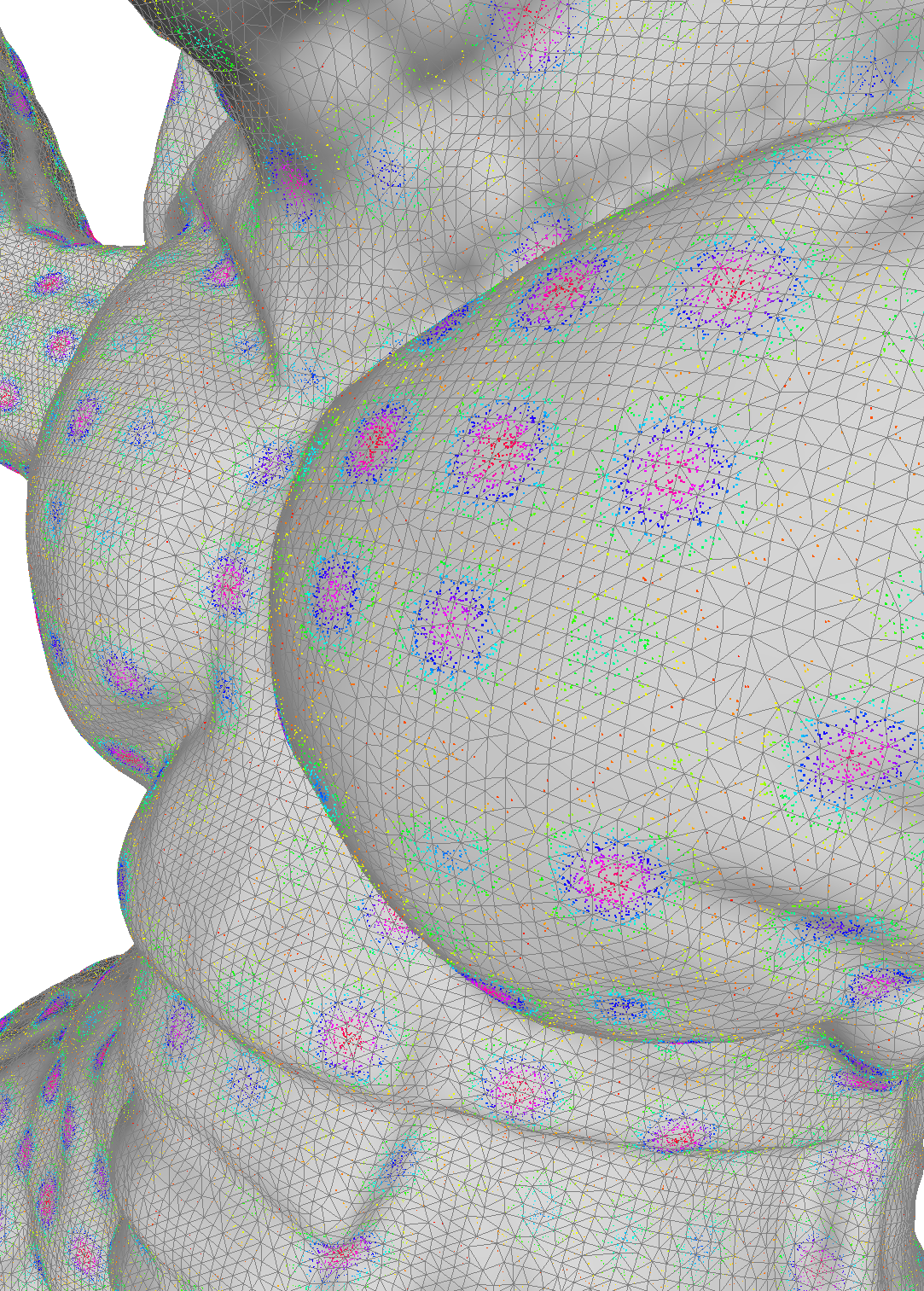}} 
\hfill
\caption{One million points distributed on a mesh with density proportional to a per-vertex periodic weight function, via inversion sampling. The point hue indicates the variation of weight magnitude from minimum to maximum. The detail on the right shows the relation of the sampled points to the underlying mesh.}
\label{fig:points_grid}
\end{figure*}

In Figure~\ref{fig:points_curv} we show a simple example of a point distribution sampled via this inversion method, in which the per-vertex weight is taken to be the magnitude of the local discrete vertex curvature. In Figure~\ref{fig:points_grid} we show another example in which the per-vertex weight function has a periodic 3d variation of the form $\left|\cos(x/L) \cos(y/L) \cos(z/L)\right|$ with some length-scale $L$. 

\begin{figure*}[!t]
\centering
\subfigure[\textbf{Kolmogorov-Smirnov test:} Plots of the CDF $P_V(v)$ from theory and from empirical sampling via the routines \textbf{\detokenize{U}} and \textbf{\detokenize{V}}. The colors indicate the region of the $(\Phi_u, \Phi_v)$ plane in Figure~\ref{weight_triangle} used for that test.\label{subfig:ks}]{\includegraphics[width=0.33\textwidth]{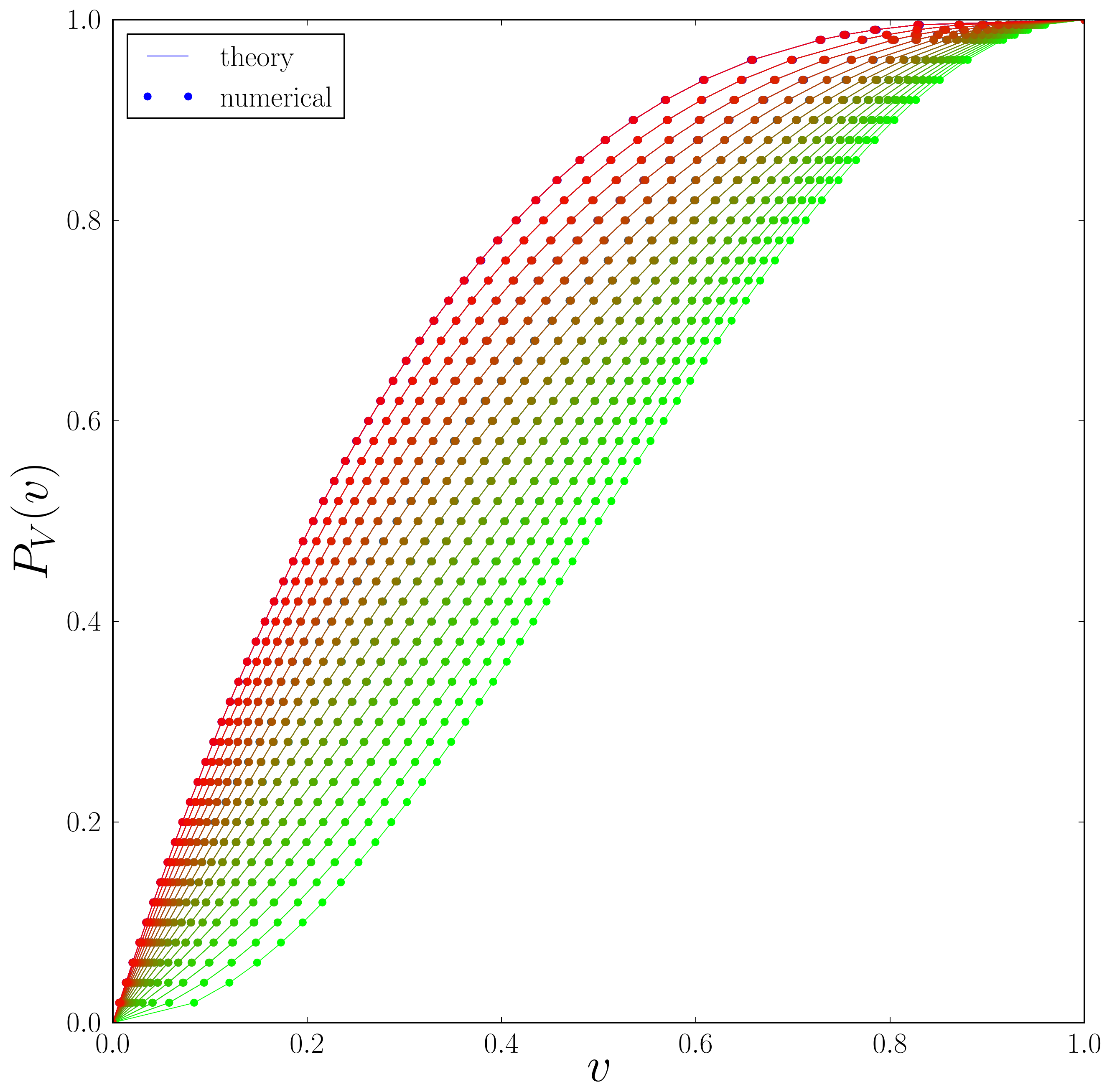}}
\hspace{15mm} 
\subfigure[\textbf{Performance:} Time per sample for point sampling in a single triangle via inversion sampling compared to rejection sampling. \label{subfig:perf}]{\includegraphics[width=0.33\textwidth]{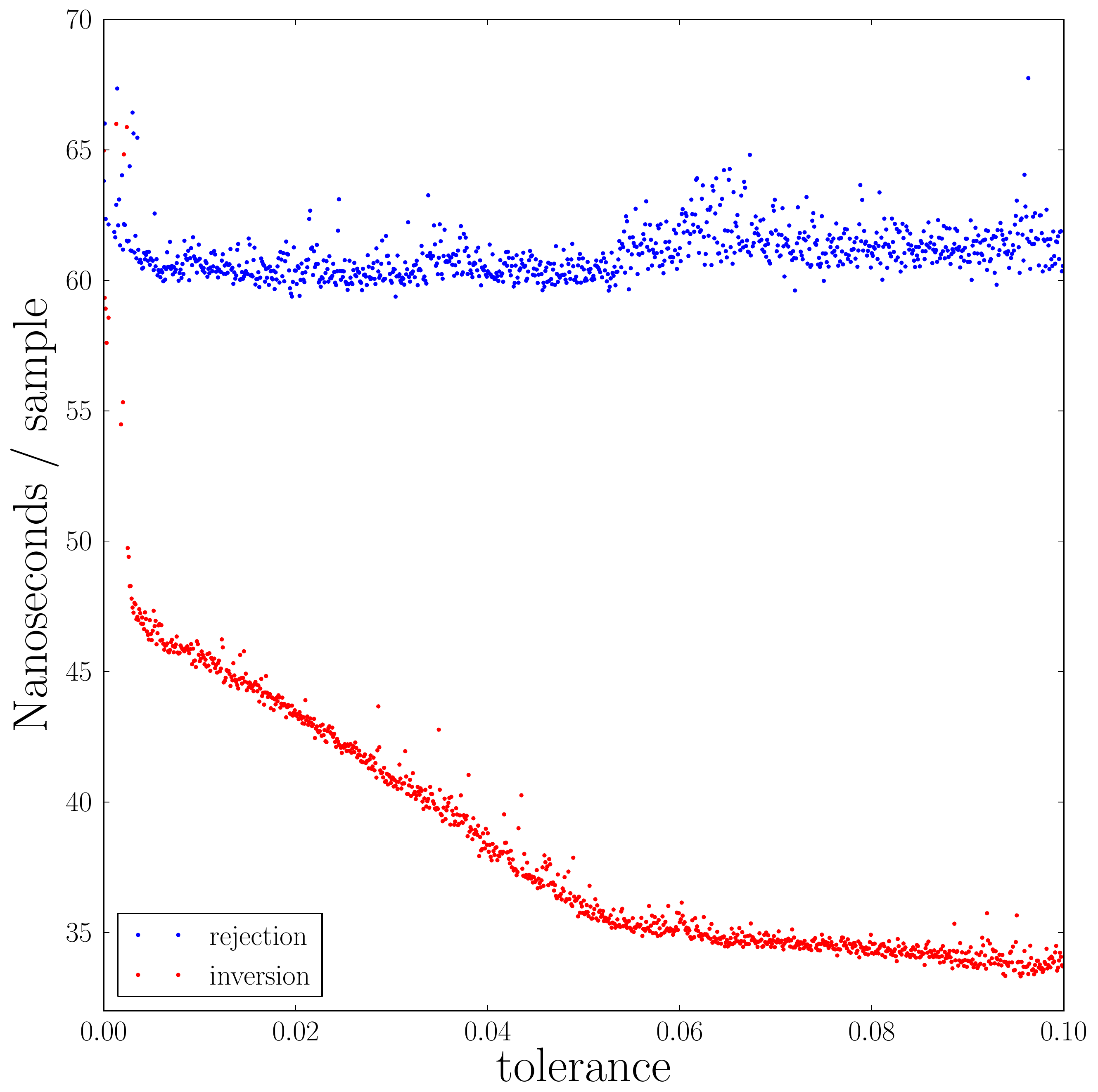}}
\caption{Validation and performance. \label{fig:validation}}
\end{figure*}

\section{Validation and Performance} \label{sec:validation}

In Figure~\ref{subfig:ks}, we plot the calculated CDF $P_V(v)$ and empirically measured CDF obtained via the inversion sampling of Algorithm~\ref{inversion}, for a representative set of triangle weightings (here those which fit in a 16x16 grid covering the valid region in the $(\Phi_u, \Phi_v)$ plane in Figure~\ref{weight_triangle}). Only 0.1\% of the CDF points are shown for clarity. 
The Kolmogorov-Smirnov statistic for each empirical CDF is at most $D=0.005$, which is less than the critical statistic at 99\% confidence level ($D_\mathrm{crit}=1.63/\sqrt{N}=0.007$),  giving confidence that the sampling algorithm is correct over the whole $(\Phi_u, \Phi_v)$ region. 

We focus here only on performance of the sampling within a given triangle, as we do not deal with optimization of the triangle selection itself. In Figure~\ref{subfig:perf} we show the relative performance of the rejection and inversion methods applied to independent samples from a single triangle (with $\Phi_u=-3$, $\Phi_v=-3$, i.e. the maximal per-triangle ``inhomogeneity'') where each data point indicates the average sample time in nanoseconds averaged over $10^7$ samples. These timings were taken running single-threaded on a \SI{2.3}{\giga\hertz} Intel Core i7 processor. The inversion method runs approximately $20$-$80$\% faster than rejection, where the efficiency depends strongly on the required tolerance for the sampled $u$ barycoordinate, so some trade-off between accuracy and speed is involved. Of course, both inversion and rejection algorithms can also be easily multi-threaded. 

\section{Conclusion}

We derived a simple and efficient inversion method for sampling points on a triangle mesh with density defined by barycentrically interpolated per-vertex weights, and verified that it produces statistically correct point samples. We showed that weighted point sampling on a triangle mesh via the inversion method is faster on average than rejection sampling.

We note that the method presented here can overall be regarded as complementary to that of Sik et al. \cite{Sik:2013:FRS}. 
Their method involves usage of a more sophisticated method for triangle sampling, and is more general as it can deal with arbitrarily varying weight functions (as can rejection sampling), however it is also considerably more complex to implement. While an algorithm such as their fast triangle selection method is required for optimal performance, using our analytical inversion sampling of linearly varying weights should allow for further improvement as it would allow less subdivision to achieve the same sampling accuracy. We suggest that in future work it would be interesting to explore the combination of these methods.

\bibliographystyle{eg-alpha-doi}
\bibliography{ewrsm}

\newcommand{\etalchar}[1]{$^{#1}$}
\begin{thebibliography}{\uppercase{YWW14}}

\bibitem[Arv01]{SS2M}
\textsc{Arvo J.}:
\newblock Stratified sampling of 2-manifolds.
\newblock \emph{SIGGRAPH 2001 Course Notes 29} (2001).

\bibitem[CCS12]{Corsini:2012:EFS:2197070.2197094}
\textsc{Corsini M., Cignoni P., Scopigno R.}:
\newblock Efficient and flexible sampling with blue noise properties of
  triangular meshes.
\newblock \emph{IEEE Transactions on Visualization and Computer Graphics 18}, 6
  (June 2012), 914--924.

\bibitem[CJW{\etalchar{*}}09]{Cline:2009:DTS:2383586.2383611}
\textsc{Cline D., Jeschke S., White K., Razdan A., Wonka P.}:
\newblock Dart throwing on surfaces.
\newblock In \emph{Proceedings of the Twentieth Eurographics Conference on
  Rendering} (2009), EGSR'09, pp.~1217--1226.

\bibitem[GP07]{Gross:2007:PG:1202384}
\textsc{Gross M., Pfister H.}:
\newblock \emph{Point-Based Graphics}.
\newblock Morgan Kaufmann Publishers Inc., San Francisco, CA, USA, 2007.

\bibitem[PH10]{Pharr:2010:PBR:1854996}
\textsc{Pharr M., Humphreys G.}:
\newblock \emph{Physically Based Rendering, Second Edition: From Theory To
  Implementation}, 2nd~ed.
\newblock Morgan Kaufmann Publishers Inc., San Francisco, CA, USA, 2010.

\bibitem[PTVF92]{Press:1992:NRC:148286}
\textsc{Press W.~H., Teukolsky S.~A., Vetterling W.~T., Flannery B.~P.}:
\newblock \emph{Numerical recipes in C (2nd ed.): the art of scientific
  computing}.
\newblock Cambridge University Press, New York, NY, USA, 1992.

\bibitem[SA07]{10.1109/RT.2007.4342601}
\textsc{Subr K., Arvo J.}:
\newblock Steerable importance sampling.
\newblock \emph{IEEE/ EG Symposium on Interactive Ray Tracing 2007 00},
  undefined (2007), 133--140.

\bibitem[{\v{S}}K13]{Sik:2013:FRS}
\textsc{{\v{S}}ik M., K{\v{r}}iv{\'a}nek J.}:
\newblock Fast random sampling of triangular meshes.
\newblock In \emph{Pacific Graphics, Short Papers} (2013).

\bibitem[ST04]{smith2004sampling}
\textsc{Smith N., Tromble R.}:
\newblock Sampling uniformly from the unit simplex.
\newblock \emph{Johns Hopkins University, Tech. Rep.} (2004), 1--6.

\bibitem[Tur90]{Turk:1990:GRP:90767.90772}
\textsc{Turk G.}:
\newblock Graphics gems.
\newblock Academic Press Professional, Inc., San Diego, CA, USA, 1990,
  ch.~Generating Random Points in Triangles, pp.~24--28.

\bibitem[YWW14]{6811174}
\textsc{Yan D.~M., Wallner J., Wonka P.}:
\newblock Unbiased sampling and meshing of isosurfaces.
\newblock \emph{IEEE Transactions on Visualization and Computer Graphics 20},
  11 (2014), 1579--1589.

\end{thebibliography}


\end{document}